\documentclass[useAMS,usenatbib]{mn2e}
\usepackage{graphicx}
\usepackage{setspace}
\usepackage{amsmath}
\usepackage{color}
\definecolor{RED}{RGB}{255,0,0}
\newcommand{\corr}{}

\title[High temperature clouds of super-hot Jupiters]{High temperature condensate clouds in super-hot Jupiter atmospheres}
\author[H. R. Wakeford et al.]{H. R. Wakeford$^{1}$\thanks{hannah.wakeford@nasa.gov} 
C. Visscher$^{2}$, 
N.K. Lewis$^{3}$, 
T. Kataria$^{4}$, 
M.S. Marley$^{5}$, \newauthor 
J.J. Fortney$^{6}$, 
and A.M. Mandell$^{1}$ \\
$^{1}$Planetary Systems Lab, NASA Goddard Space Flight Center, Greenbelt, MD 20771, USA\\
$^{2}$ Department of Chemistry, Dordt College, Sioux Center, Iowa 51250, USA\\
$^{3}$ Space Telescope Science Institute, 3700 San Martin Drive, Baltimore, MD 21218, USA\\
$^{4}$ NASA Jet Propulsion Laboratory, 4800 Oak Grove Dr, Pasadena, CA 91109, USA\\
$^{5}$ NASA Ames Research Center, MS 245-5, Moffett Field, CA 94035, USA\\
$^{6}$ Department of Astronomy \& Astrophysics, University of California, Santa Cruz, CA 95064, USA}

\begin{document}

\date{Accepted 2016 October 11.}
%\date{Aug 2016}

\pagerange{\pageref{firstpage}--\pageref{lastpage}} \pubyear{2016}

\maketitle

\label{firstpage}

\begin{abstract}
Deciphering the role of clouds is central to our understanding of exoplanet atmospheres, as they have a direct impact on the temperature and pressure structure, and observational properties of the planet. Super-hot Jupiters occupy a temperature regime similar to low mass M-dwarfs, \corr{where minimal cloud condensation is expected.} However, observations of exoplanets such as WASP-12b (T$_{eq}\sim$2500\,K) result in a transmission spectrum indicative of a cloudy atmosphere. We re-examine the temperature and pressure space occupied by these super-hot Jupiter atmospheres, to explore the role of the initial Al- and Ti-bearing condensates as the main source of cloud material. 
Due to the high temperatures a majority of the more common refractory material is not depleted into deeper layers and would remain in the vapor phase. 
The lack of depletion into deeper layers means that these materials with relatively low cloud masses can become significant absorbers in the upper atmosphere. We provide condensation curves for the initial Al- and Ti-bearing condensates that may be used to provide quantitative estimates of the effect of metallicity on cloud masses, as planets with metal-rich hosts potentially form more opaque clouds because more mass is available for condensation. Increased metallicity also pushes the point of condensation to hotter, deeper layers in the planetary atmosphere further increasing the density of the cloud. We suggest that planets around metal-rich hosts are more likely to have thick refractory clouds, and discuss the implication on the observed spectra of WASP-12b.
\end{abstract}


\section{Introduction}
Since the discovery of hot Jupiters, Jupiter-sized planets that orbit close to their host stars (\textless0.1\,AU), these planets have been subject to intensive observational and theoretical campaigns to characterize the nature of this uncharted planetary regime (e.g. \citealt{loddersfegley2002}, \citealt{seager2010book}, \citealt{fortney2010}, \citealt{haswell2012}, \citealt{burrows2014}, \citealt{showman2015}, \citealt{Sing2016}). Out of the nearly 2,000 exoplanets confirmed thus far, over 50 are expected to have equilibrium temperatures (T$_{eq}$) greater than 1800\,K, assuming full heat redistribution around the planet, with a majority falling into the Jupiter mass regime. These ``super-hot'' Jupiters often have ultra short period orbits (less than four days) around high metallicity G and F-type stars. As a result of their high temperatures, super-hot Jupiters likely have atmospheric compositions similar to late-M stars (T$\approx$2245--1960\,K) and L dwarfs (T\textless1950\,K). 

Similar to M stars, \corr{giant} exoplanets with incident flux levels \textgreater10$^9$\,ergs$^{-1}$cm$^{-2}$ (\corr{approximately} \textless0.04\,AU) were thought to be cloud-free with gaseous TiO and VO contributing as a major opacity source in the atmosphere, which results in thermal inversions on the dayside (e.g. \citealt{hubeny2003}, \citealt{fortney2008}). This suggests a vast majority of super-hot Jupiters fall into this category where only vapor phase products will be observed in both transmission and emission. Out of the known super-hot Jupiters only a handful have been studied in detail through observations. Currently only one planet, WASP-33b (T$_{eq}$=2710\,K), shows evidence for a thermal inversion in the dayside emission, however, evidence for TiO emission features in the observations remains inconclusive (\citealt{haynes2015}). Additional, tentative evidence for TiO in the transmission spectrum of WASP-121b (T$_{eq}$=2400\,K) has been presented in \citet{evans2016}, but confirmation is still needed from higher resolution space-based data. Transmission spectral studies of WASP-19b (T$_{eq}$=2077\,K) and WASP-12b (T$_{eq}$=2580\,K), two super-hot Jupiters observed as part of the large transmission spectral survey presented in \citet{Sing2016}, show no evidence for TiO/VO absorption suggesting it has been removed from the atmosphere. One possibility of this discrepancy is that the TiO is being cold-trapped on the planets nightside due to the planets atmospheric circulation through horizontal and vertical mixing \corr{(e.g. \citealt{spiegel2009}; \citealt{parmentier2013})}. However, observations of the hotter WASP-12b suggest the presence of a substantial cloud deck obscuring additional atomic and molecular features in the atmosphere at the dayside/nightside transition. Despite its high temperature, this appears more typical of the transition from M- to L-dwarf atmospheres \corr{(e.g. \citealt{allard2000,allard2001,allard2011})}, suggesting that material is being sustained as liquid or solids at optically thick volumes high in the atmosphere. 

The presence or absence of clouds in the atmosphere has strong implications on the total energy budget of the the planet, as they can have a large affect on the absorption properties of the observed photosphere (\citealt{sanchez2004}). The presence of clouds in a planetary atmosphere can remove absorbers from the gas phase, obscure absorption features from gases at deeper levels (\citealt{Sing2016}), scatter incoming radiation, and add their own absorption features (\citealt{wakeford2015}). A number of approaches have been used to understand the formation of condensate clouds in substellar atmospheres. Often these scenarios require nucleation and the presence of seed particles on which the material can condense and be transported through the atmosphere. In one scenario cloud formation occurs through sedementation and upward mixing of seed particles through the atmosphere (e.g. \citealt{ackerman2001}; \citealt{loddersfegley2006}). A second scenario uses a top-down approach accounting for micro-physical processes to grow particles and they settle through the atmosphere assuming vertical mixing timescales are faster than the condensation timescales (e.g. \citealt{Helling2008cloud}, \citeyear{Helling2009a}, \citeyear{Helling2009b}). A comprehensive review of different cloud modeling techniques and predictions can be found in \citet{Helling2008review}. Many of these models are developed from brown dwarf studies where more comparative data is available to constrain the model parameter space, however, discrepancies still occur. As more and more exoplanet data becomes available through ground and space-based observatories, such as the James Webb Space Telescope (JWST), large grids of models will be needed to make comparisons and predictions about cloud formation and species in exoplanet atmospheres. In addition to this, laboratory-based studies will be required to fill in the optical properties of different condensate species expected at higher temperatures and lower pressures in non-N$_2$ based atmospheres (\citealt{fortney2016}).

In an atmosphere bound by gravity, settling of condensate species will likely form discrete cloud layers (\citealt{marley1999b}, \citealt{loddersfegley2006}, \citealt{visscher2010}), which can change the reaction chemistry at that temperature and pressure (\citealt{marley2007}). In the atmospheres of late M-stars dust likely forms in clouds evenly distributed around the surface at the deepest/hottest level where the grain type can condense (\citealt{allard2001}). Typical M/L dwarf atmospheric temperature-pressure (T-P) profiles are shallow, crossing multiple condensation curves where clouds can form across a wide temperature range. The strong external heating of super-hot Jupiters compared to lone brown dwarfs and M-stars results in much steeper T-P profiles in the upper atmosphere. This causes the upper atmospheric T-P profile to run parallel to multiple condensation curves, where small changes in local conditions can shift the T-P profile across condensation lines making their atmosphere very sensitive to cloud formation (\citealt{Sing2016}). The tidally locked nature of these worlds also results in vast differences in temperature from the dayside to the nightside which can have an effect on the clouds and chemistry in different regions of the atmosphere (\citealt{kataria2016}). 

Here we postulate that the same cloud species found in the M-L transition are important for WASP-12b and other super-hot Jupiters and suggest the phase space where these clouds will have the strongest observational impact. We explore the high-temperature condensate sequence likely to occur in super-hot Jupiter atmospheres, examine the impact of metallicity on atmospheric chemistry where the equilibrium temperature exceeds 1800\,K, and discuss the effect this has on the observational properties of condensates in the atmosphere. In \S\ref{sec:w12} we examine the case of WASP-12b and discuss the potential cloud condensates responsible for the observed transmission spectrum.\\

\begin{figure*}
\centering 
  \includegraphics[width=0.99\textwidth]{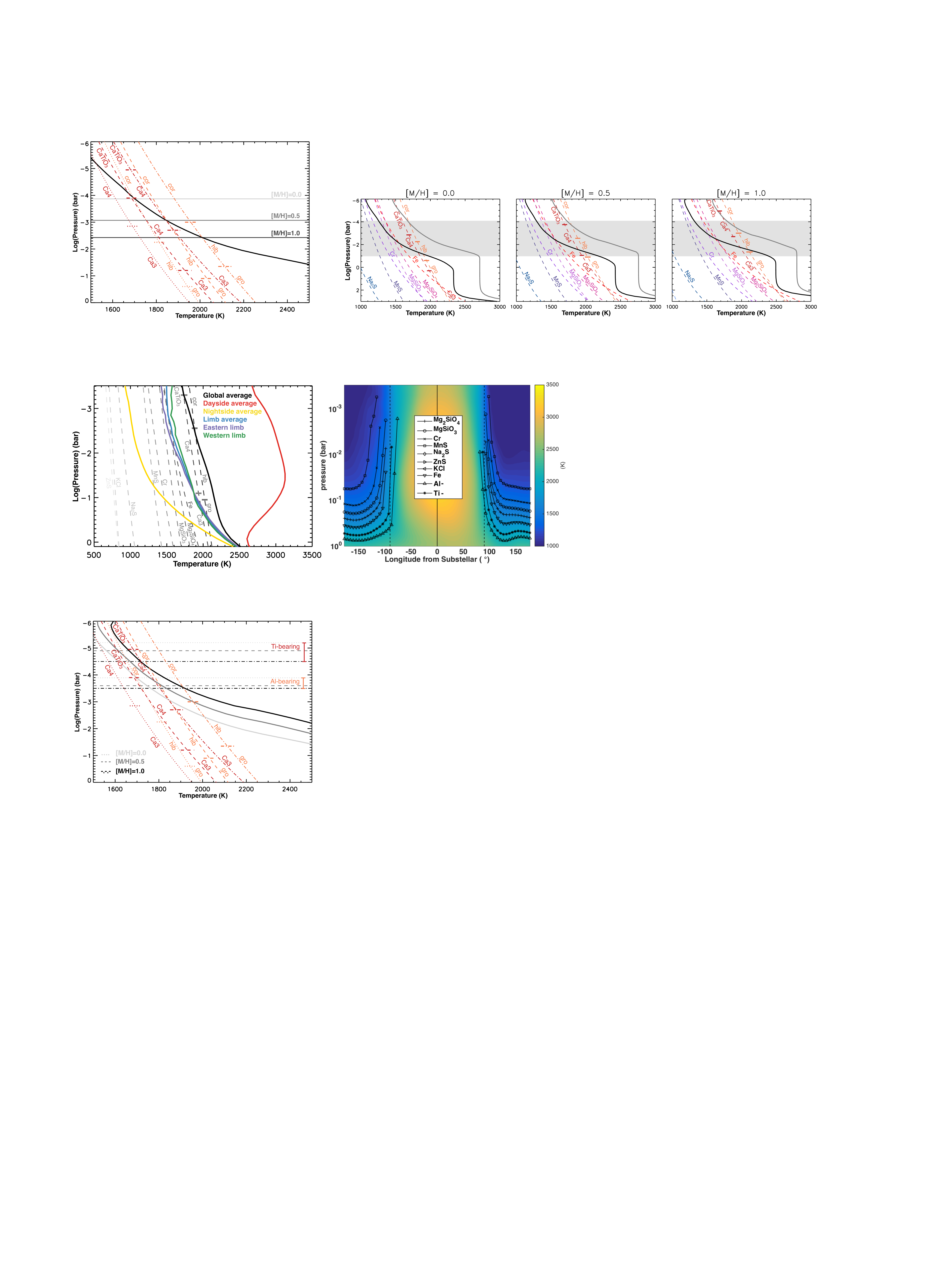}%shj_tp_v4 with Fig 2 box
\caption{Each panel displays the 1D T-P profile \corr{(\citealt{fortney2008})} of typical super-hot Jupiters orbiting a G0V star with T$_{eq}\approx$1860\,K (left profile) and T$_{eq}\approx$2500\,K (right profile), g$_p$\,=\,10\,ms$^{-2}$, and T$_{int}$\,=\,100\,K. These have been computed for [M/H]\,=\,0.0, 0.5, 1.0, and plotted against the corresponding condensation curves. Condensation curves are computed for Al- and Ti-bearing species following the equations outlined in \citet{visscher2010} and Section\,\ref{sec:metallicity}. Following the convention of \citet{lodders2002}, abbreviated condensate names are \emph{gro} = grossite (CaAl$_{4}$O$_{7}$), \emph{hib} = hibonite (CaAl$_{12}$O$_{19}$), \emph{cor} = corundum (Al$_2$O$_3$), \emph{Ca3} = Ca titanate Ca$_3$Ti$_2$O$_7$, and \emph{Ca4} = Ca titanate Ca$_4$Ti$_3$O$_{10}$. The short horizontal lines on the Al- and Ti- bearing condensation curves show the point of initial condensation, with each portion of the curve labeled for the \corr{initial Al-bearing or Ti-bearing condensate, respectively.}
The grey shaded regions show the pressures observed with transmission spectroscopy.}\label{fig:shj_pt_profiles}
\end{figure*}% will you look at that caption, wowza!

\begin{table}
 \caption{Relative Cloud Masses and Condensation Temperatures for $P_T=0.1$ bar and [Fe/H] = 0.0, after \citet{lodders2002}.}\label{table:cloud_mass}
 \label{tab:example}
 \begin{tabular}{llcc}
  \hline
  Name  & Formula & Relative & Condensation\\
 & & Cloud Mass$^\dagger$ & Temperature (K)\\
  \hline
\multicolumn{4}{c}{\emph{Initial Al-bearing condensates}}\\
Hibonite  		& CaAl$_{12}$O$_{19}$   				& 2091   			& {1914}		\\
Grossite 		& CaAl$_{4}$O$_{7}$   				& 2442   			& 1848$^{\ddagger}$\\
Corundum  	& Al$_{2}$O$_{3}$   					& 1915   			& ...		  		\\

\multicolumn{4}{c}{\emph{Initial Ti-bearing condensates}}\\
Ca titanate  	& Ca$_{4}$Ti$_{3}$O$_{10}$   			& 170  			& 1854   		\\
Ca titanate 	& Ca$_{3}$Ti$_{2}$O$_{7}$   			& 180   			& 1818$^{\ddagger}$     	\\
Perovskite 	& CaTiO$_{3}$  					& 149   			& 1707$^{\ddagger}$		\\
\multicolumn{4}{c}{\emph{Other condensates}}\\
Gehlenite 		& Ca$_{2}$Al$_{2}$SiO$_{7}$			& 3678   			& 1796$^{\ddagger}$   		\\
Fe metal 		& Fe   							& 21032  			& 1682   		\\
Spinel 		& MgAl$_{2}$O$_{4}$  				& 2673   			& 1663$^{\ddagger}$   		\\
Akermanite 	& Ca$_{2}$Mg$_{2}$Si$_{2}$O$_{7}$	& 3657			& 1626$^{\ddagger}$  		\\
Ti oxide 		& Ti$_{3}$O$_{5}$  					& 82   			& 1610$^{\ddagger}$  		\\
Ti oxide 		& Ti$_{2}$O$_{3}$  					& 79   			& 1610$^{\ddagger}$  	\\
Forsterite 		& Mg$_{2}$SiO$_{4}$  				& 32180  			& 1592   	\\
Diopside 		& CaMgSi$_{2}$O$_{6}$  				& 5809  			& 1560$^{\ddagger}$    	\\
Anorthite 		& CaAl$_{2}$Si$_{2}$O$_{8}$  			& 5227  			& 1513$^{\ddagger}$    	\\
Enstatite 		& MgSiO$_{3}$  					& 44585  			& 1508    			\\
Cr metal 		& Cr 								& 303 			& 1416$^{\ddagger}$		\\
Ti oxide 		& Ti$_{4}$O$_{7}$  					& 83   			& 1290$^{\ddagger}$  		\\
Mn sulfide 	& MnS 							& 360			& 1271			\\
Na sulfide 	& Na$_{2}$S 						& $\equiv$ 1000	& 925		\\
  \hline
 \end{tabular}
$^\dagger$Relative cloud masses are normalized to Na$_{2}$S $\equiv$1000, assuming complete condensation of a given condensate using bulk atmospheric element inventories and element abundances from \citet{lodders2010}.  See text for details.\\
$\ddagger$First appearance in the condensation sequence at 0.1 bar (temperatures from \citealt{lodders2002}).  Al$_2$O$_3$ does not appear in the sequence at this pressure, however is expected at lower pressures (cf. \citealt{lodders2002}, \citealt{ebel2006}).  The removal of material by formation of the initial, higher-temperature condensates influences the behavior of subsequent (lower-temperature) Ca-, Al-, and Ti-bearing condensates. 
\end{table} %Please excuse my many notes it is a matter of overcompensation, and a smidge of fear.

\section{High-temperature cloud condensates} \label{sec:clouds}
Ca, Ti, and Al have been shown to form the highest-temperature condensates at atmospheric pressures relevant to M-stars and L-type brown dwarfs (\citealt{burrows1999}, \citealt{lodders1999}, \citealt{allard2001}, \citealt{lodders2002}). The stability fields of condensates formed from these elements under equilibrium conditions span pressures from 100s of bars to microbars and temperatures from $\sim$2500 to 1500\,K, and are intricately linked to the gas chemistry of the atmosphere. Here we revisit earlier studies of high-temperature condensates, in particular the condensation sequence of \citet{lodders2002}, and apply these earlier results to explore the role of cloud formation in super-hot Jupiters such as WASP-12b.

We calculate gas and condensate chemistry following the detailed thermochemical models of previous studies (e.g., \citealt{fegley1994}, \citeyear{fegley1996}, \citealt{burrows1999}, \citealt{lodders1999}, \citeyear{lodders2002}, \citeyear{lodders2010}, \citealt{loddersfegley2002}, \citeyear{loddersfegley2006}, \citealt{visscher2006}, \citeyear{visscher2010}, \citealt{morley2012}). We assume equilibrium cloud formation occurring upon vapor pressure saturation via vertical mixing in an atmospheric column, to primarily determine the composition and total quantity of condensible material. This equilibrium cloud condensation scenario (initially developed by Lewis 1969 for Jupiter) is supported by several lines of evidence in model-data comparisons of exoplanet and brown dwarf atmospheres (e.g., see \citealt{loddersfegley2006}, \citealt{visscher2010}, and references therein). Here we focus on the role of the highest-temperature Al-bearing condensates grossite (CaAl$_4$O$_7$), hibonite (CaAl$_{12}$O$_{19}$), and corundum (Al$_2$O$_3$) and Ti-bearing condensates Ca$_3$Ti$_2$O$_7$, Ca$_4$Ti$_3$O$_{10}$, and perovskite (CaTiO$_{3}$). These represent the initial (i.e., highest-temperature) Al- and Ti-bearing condensates, respectively, and may be used to estimate the total mass of refractory cloud material expected in observable regions of super-hot Jupiter atmospheres.

\subsection{Condensate cloud masses} 
To understand the impact a condensate will have on the observed properties of a planetary atmosphere, we first have to determine the amount of material available to condense into a cloud. Table\,\ref{table:cloud_mass} shows the condensation temperatures and relative cloud masses using elemental abundances from \citet{lodders2010} and assuming the maximum possible mass for each condensate (normalized to Na$_{2}$S $\equiv$ 1000; see \citealt{morley2012}). In this approach, we assume complete removal of available condensible material from the gas phase, using the total atmospheric inventory of each element (cf. \citealt{burrows1999}, \citealt{marley2000}, \citealt{morley2012}). The maximum possible cloud mass is thus limited by the least abundant species in each condensate. The relative mass ($m_i'$) of a condensate $i$ in Table\,\ref{table:cloud_mass} can be expressed as
\begin{equation}
m_i' = \frac{(A_M/n_{M(i)})\mu_i }{(A_{\textrm{Na}}/n_{\textrm{Na}\textrm{(Na}_2\textrm{S)}})\mu_{\textrm{Na}_2\textrm{S}}}\times1000
\end{equation}
where $A_M$ is the atomic abundance of the limiting element $M$ (from \citealt{lodders2010}), $n_{M(i)}$ is the stoichiometric number of atoms of element $M$ in one molecule of $i$ ($n_{\textrm{Na}\textrm{(Na}_2\textrm{S)}}=2$), and $\mu_i$ is the molecular weight of condensate $i$. The relative cloud masses provided in Table\,\ref{table:cloud_mass} are thus calculated assuming complete removal of the limiting element by each condensate. This approach provides an upper limit on the mass of each cloud condensate. 

For example, the complete condensation of hibonite (CaAl$_{12}$O$_{19}$) would consume nearly all atmospheric Al but just $\sim12\%$ of the Ca inventory, such that the maximum theoretical relative mass of hibonite is limited by the total Al inventory of the atmosphere (given the 12:1 Al:Ca ratio in hibonite). The complete condensation of grossite (instead of hibonite or corundum) or corundum (instead of hibonite or grossite) would likewise consume nearly all atmospheric Al. The maximum cloud mass is thus roughly similar for each of the initial Al-bearing condensates (see Table\,\ref{table:cloud_mass}), as the mass of each of these condensates is limited by the total Al inventory of the atmosphere. On the other hand, the complete condensation of gehlenite (Ca$_2$Al$_2$SiO$_7$) at slightly lower temperatures  would theoretically consume nearly all atmospheric Ca and 71\% of the Al inventory, as gehlenite possesses a 1:1 Ca:Al ratio and Al is 1.4$\times$ more abundant than Ca in a solar-composition gas (\citealt{lodders2010}). In the same way, the maximum possible mass of the highest-temperature Ti-bearing condensate Ca$_3$Ti$_2$O$_7$, Ca$_4$Ti$_3$O$_{10}$, or perovskite (CaTiO$_{3}$) is limited by the total inventory of Ti in the atmosphere.

We note that in contrast to Table\,\ref{table:cloud_mass}, several elements are likely to be distributed among more than one condensed phase in the atmosphere (e.g., Mg, Si into forsterite and enstatite; Ca, Al, Ti into several possible phases; see \citealt{burrows1999}, \citealt{lodders2002}, \citealt{visscher2010}; see \citealt{lodders2002} for a detailed treatment of equilibria between condensed phases) and the removal of elemental material by the initial (higher-temperature) condensates will influence the condensation behavior of subsequent (lower-temperature) phases. Here we estimate maximum theoretical cloud masses in order to estimate the quantity of material expected to condense in the upper atmospheres of super-hot Jupiters.  

The condensation sequence will also vary with changes in pressure (\corr{\citealt{lodders2002}}).  For example, the initial Al-bearing condensate is grossite (CaAl$_4$O$_7$) at high pressures, hibonite (CaAl$_{12}$O$_{19}$) at intermediate pressures, and corundum (Al$_2$O$_3$) at low pressures in a solar-composition gas (see \corr{Section\,\ref{sec:metallicity}}; cf. \citealt{lodders2002}, \citealt{ebel2006}).  Similarly, the initial Ti-bearing condensate is Ca$_3$Ti$_2$O$_7$ at high pressures, Ca$_4$Ti$_3$O$_{10}$ at intermediate pressures, and perovskite (CaTiO$_{3}$) at low pressures in a solar-composition gas (cf. \citealt{burrows1999}, \citealt{lodders2002}).

\begin{figure} 
\begin{center}
  \includegraphics[width=0.49\textwidth]{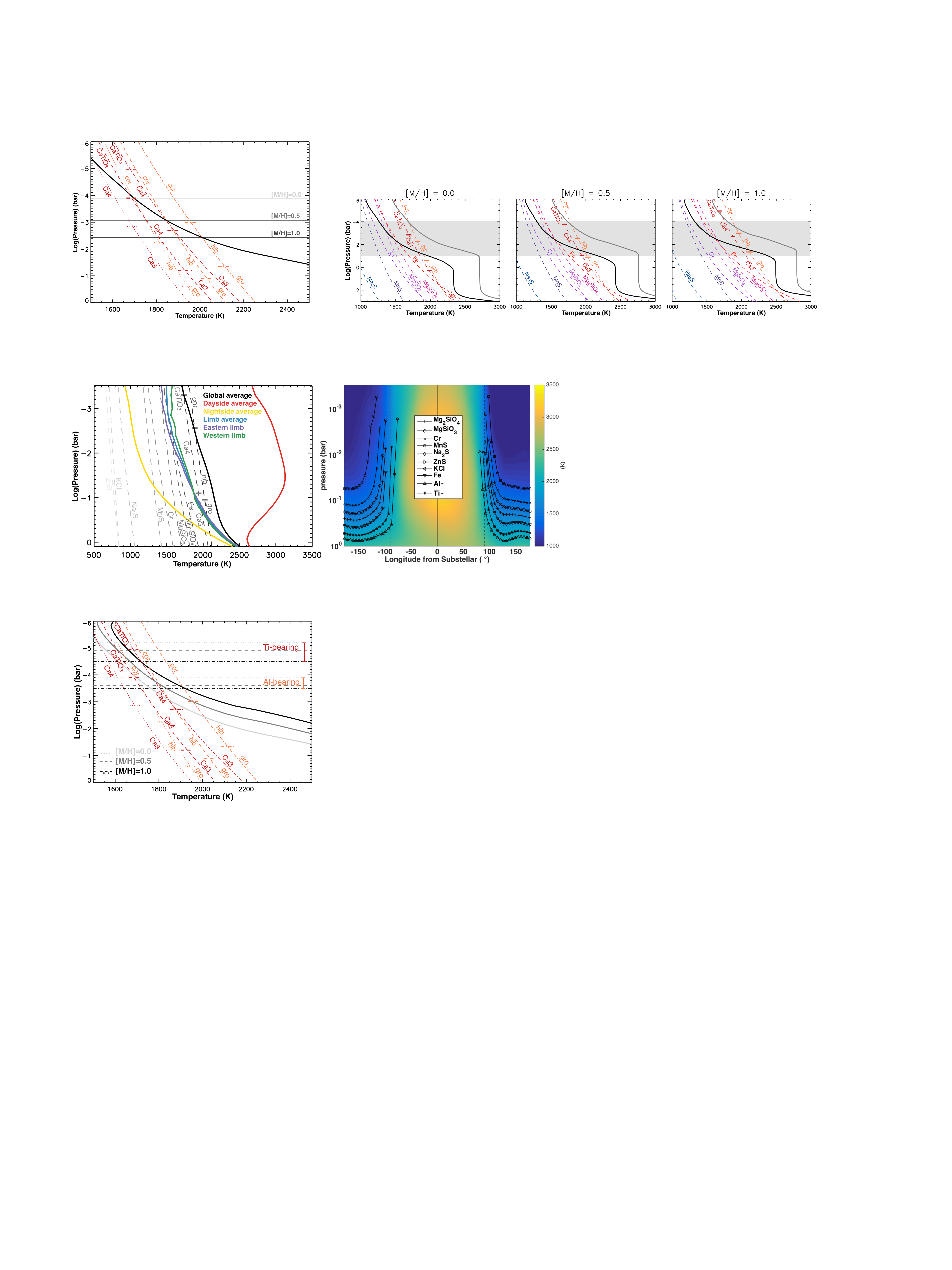}
\caption{T-P profile of a super-hot Jupiter with T$_{eq}\approx$2500\,K. Same as Fig.\,\ref{fig:shj_pt_profiles} the grey lines show the expected cloud base pressure for varied metallicity atmospheres, with the corresponding condensation curves for grossite, hibonite, and corundum at [M/H]=0.0 (dotted), 0.5 (dashed), and 1.0 (dash-dot) with the colored horizontal lines on the curves showing the point of initial condensation of each species.} \label{fig:shj_cloud_base} 
\end{center}
\end{figure}

\section{Effects of Metallicity}\label{sec:metallicity}
The location and composition of clouds observed in planetary atmospheres under equilibrium conditions can be estimated by observing the point at which the condensation curve of a particular species crosses the model T-P profile of the atmosphere. The condensation curves themselves are independent of the planetary T-P profile and vary based on pressure, temperature, and metallicity, but do depend on the abundances. Therefore, any specific T-P profile for different regions of the planetary atmosphere can be superimposed to determine the planetary equilibrium chemistry. Note, we do not consider the thermal feedbacks likely associated with the cloud formation but instead overlay the atmospheric profiles with the condensation curves in order to approximate the condensation temperatures and cloud-base pressures for numerous objects simultaneously. \corr{As was shown in \citet{morley2012} feedback from cloud has only minor impact on the T-P profiles and do not change the overall trends observed or results.}

\corr{We use planetary-averaged T-P profiles of two super-hot Jupiters, computed following the methods of \citet{fortney2008} for a cloud-free atmosphere, as examples of a cooler atmosphere T$_{eq}\approx$1860\,K and a hotter atmosphere T$_{eq}\approx$2500\,K (see Fig.\,\ref{fig:shj_pt_profiles}). These profiles assume a H/He dominated giant planet with g$_p$\,=\,10\,ms$^{-2}$, orbiting a G0V star (L\,=\,1.35\,L$\odot$), and an assumed internal temperature of 100\,K. The internal temperature used is} a reasonable estimate given the range of ages and surface gravities estimated for the hot Jupiter population \citep[e.g.][]{fortney2007} and consistent with the intrinsic luminosities of Jupiter and Saturn \citep{hubbard1980}. 

\corr{The impact of changes in metallicity can be seen both on the T-P profile and the condensation curves (see Fig.\,\ref{fig:shj_pt_profiles} and \ref{fig:shj_cloud_base}).} We show the effect of increasing metallicity on the position of condensation in the planetary atmosphere for [M/H] = 0.0 (solar), 0.5, and 1.0. In the different metallicity scenarios presented in Fig.\,\ref{fig:shj_pt_profiles}, the first cloud forming layers appear in the observable region for transmission spectral studies, indicated by the gray shaded region.
% CHECK here
\corr{In regions of the atmosphere where the T-P profile tends toward isothermal, increasing the metallicity has the effect of shifting the cloud base pressures to hotter, deeper layers in the planetary atmosphere (see Fig. \ref{fig:shj_cloud_base}). As the profiles tend toward isobaric the temperature of the atmosphere increases with that of the condensate, therefore the pressure at which the cloud forms are not impacted as strongly. The condensation sequence is often non-linear with metallicity (see equations \ref{eqn:grossite}--\ref{eqn:perovskite}). For example the initial condensation of corundum and perovskite occur at lower pressures with increasing metallicity, while in these cases the cloud forming species is not changed, extreme increases in metallicity could result in different Al- and Ti- species becoming the dominant absorber.}
\corr{We indicate the base pressure of the Al- and Ti-bearing condensates for a specific T-P profile at different metallicities in Fig.\,\ref{fig:shj_cloud_base}, which shows the pressure difference with metallicity across different regions of the atmospheric T-P profile. The change in derived pressure regions or absorbing species from exoplanet transmission spectra may be a potential indicator for atmospheric metallicity as well as potentially indicating specific cloud species where pressure levels are well defined in observations.} 

\subsection{Initial condensation curves}
Given their similar condensation temperatures and cloud masses (see Table\,\ref{table:cloud_mass}) it may be difficult to distinguish which Al- or Ti-bearing species are being observed. We thus focus on the behavior of these two groups of condensates: the initial (highest-temperature) Al-bearing condensates (grossite, hibonite, corundum), which condense at similar temperatures (see Figure\,\ref{fig:shj_cloud_base}) and have similar relative cloud masses limited by the Al abundance, and the initial (highest-temperature) Ti-bearing condensates  (Ca titanates and perovskite), which form in the same region and have similar relative cloud masses limited by the Ti abundance.  Moreover, even considering equilibrium condensation without sedimentation (see \citealt{ackerman2001}), the initial condensates are very effective at removing condensible material from the gas phase such that nearly all Al and Ti are removed from the atmosphere within a scale height of their respective initial condensate cloud bases.  The condensation behavior of the initial condensates is thus used to approximate the amount of cloud material expected to form in the upper atmospheres of super-hot Jupiters.

Following the approach of \citet{visscher2010} for some of the phases identified in previous studies (\citealt{burrows1999}, \citealt{allard2001}, \citealt{lodders2002}), the calculated equilibrium condensation temperatures as a function of pressure and metallicity for the highest-temperature Al-bearing condensates are approximated by
\begin{multline}
    10^{4}/T_{\textrm{cond}}({\textrm{CaAl}_{4}\textrm{O}_{7}})\approx 4.990 - 0.2394 \log P_T +   \\ 1.398\times10^{-3}(\log P_T)^2-0.595[\textrm{Fe/H}], \label{eqn:grossite}
\end{multline}

\begin{multline}
      10^{4}/T_{\textrm{cond}}({\textrm{CaAl}_{12}\textrm{O}_{19}})\approx 4.997 - 0.2262 \log P_T +   \\ 1.888\times10^{-3}(\log P_T)^2-0.585[\textrm{Fe/H}], \label{eqn:hibonite}
\end{multline}

\begin{multline}
    10^{4}/T_{\textrm{cond}}({\textrm{Al}_{2}\textrm{O}_{3}})\approx 5.014 - 0.2179 \log P_T +   \\ 2.264\times10^{-3}(\log P_T)^2-0.580[\textrm{Fe/H}], \label{eqn:corundum}
\end{multline}
for $P_T$ in bars.  The equilibrium condensation temperatures for the highest-temperature Ti-bearing condensates are approximated by
\begin{multline}
     10^{4}/T_{\textrm{cond}}({\textrm{Ca}_{3}\textrm{Ti}_{2}\textrm{O}_{7}})\approx    \\ 5.114 - 0.284 \log P_T - 0.568[\textrm{Fe/H}],  \label{eqn:catitinate2}   
\end{multline}

\begin{multline}
    10^{4}/T_{\textrm{cond}}({\textrm{Ca}_{4}\textrm{Ti}_{3}\textrm{O}_{10}})\approx  \\ 5.113 - 0.281 \log P_T - 0.563[\textrm{Fe/H}], \label{eqn:catitinate}
\end{multline}

\begin{multline}
    10^{4}/T_{\textrm{cond}}({\textrm{Ca}\textrm{Ti}\textrm{O}_{3}})\approx    \\ 5.125 - 0.277 \log P_T - 0.554[\textrm{Fe/H}]. \label{eqn:perovskite}
\end{multline}

We note that the condensation temperature for each phase is estimated assuming initial condensation from an undepleted gas phase, in the absence of any prior (i.e., deeper) removal of material by cloud formation (see \citealt{lodders2002} for a detailed description of equilibria between condensed phases). Thus the above condensation curves are valid only over pressures where the Al- or Ti-bearing phase under consideration is the highest-temperature Al-bearing or Ti-bearing condensate, respectively.  For example, the grossite (CaAl$_4$O$_7$) curve is valid at high pressures ($P>1$ bar) and the corundum (Al$_{2}$O$_{3}$) curve is valid at low pressures ($P<0.1$ bar), whereas hibonte (CaAl$_{12}$O$_{19}$) is the first Al-bearing phase to condense near $P\sim0.1$ bar. In each case, the formation of the Al-bearing condensate will effectively remove Al from the vapor phase and into cloud material.  The extra pressure term for the Al-bearing condensate curves helps to account for Al-AlOH gas equilibria near cloud formation temperatures.

\subsection{Impact on cloud mass}
\corr{The maximum total mass of a given condensate will always be limited by some limiting element, the abundance of which scales directly with metallicity. For example, in corundum (Al$_2$O$_3$) Al is the limiting element, as oxygen is readily available. Therefore the relative cloud mass of corundum scales linearly with the atmospheric Al-abundance as determined by the metallicity. Using corundum to demonstrate the increase in cloud mass with increased metallicity we calculate the mass of material relative to Na$_{2}$S at solar metallicity (see Table\ref{table:cloud_mass}) at [M/H]\,=\,0.3, 0.5 and 1.0. As is shown in Table\,\ref{table:cloud_mass}, at solar metallicity corundum has a relative cloud mass of 1915 \corr{normalized to Na$_{2}$S $\equiv$1000}. When the metallicity is increased to the stated values the mass increases to 3822, 6057, and 19155 respectively. }

Assuming the cloud is condensed into \corr{a set altitude range in the atmosphere,} the spherical shell volume of the planet can be estimated. \corr{We assume the particles are well mixed throughout the extent of the cloud with a constant number density per unit volume.} The optical depth can then be approximated as a function of wavelength for each cloud condensate for a given altitude in a planets atmosphere. Therefore, by increasing the mass of the cloud the volume density, and therefore the optical depth increases. This suggests that planets orbiting metal-rich stars or with higher intrinsic metallicity are more likely to form more massive clouds. \corr{This in turn cloud produce flat or muted spectral features at altitudes observed in transmission. We expand on this with reference to super-hot Jupiters in \S\ref{sec:observables}.} 

\corr{In summary, an increase in the metallicity has a two-fold effect on the planetary atmosphere. One, in regions of the atmosphere where the T-P profile is near-isothermal the point of condensation will be shifted to deeper layers in the atmosphere, which has the effect of moving the opacity source to potentially observable pressures, as well as reducing the spherical shell volume over which the condensed material will become an opacity source. Two, critically, increasing the metallicity will increase the mass of material available to form more optically thick clouds, provided the formation pressure remains in the observable region of the atmosphere. For super-hot Jupiters this will likely result in an increased cloud opacity by Al- and/or Ti-bearing condensate species in the atmospheric transmission spectra.}

\section{Observables}\label{sec:observables}
At the high-temperature end of the condensate sequence \corr{(T$_{absolute}$\textgreater1700\,K for specific regions of the atmosphere}) it is not expected that additional refractory material \corr{(Mg, Si, Fe)} is depleted into deeper layers of the planetary atmosphere (\citealt{visscher2010}), meaning all expected material is available to condense into clouds. At temperatures greater than 1800\,K it was expected that TiO gas would dominate the optical spectra (e.g. \citealt{fortney2008}), yet current observational evidence does not support this. One possibility is that the TiO is being depleted via physical or chemical processes, and/or that another species is forming an obscuring cloud in the upper atmosphere. 

The process of forming Ti-bearing condensates in the atmosphere where temperatures fall below 1860\,K has the potential to significantly deplete TiO in the gas phase, thus reducing the opacity of TiO gas and reducing its influence on the observed transmission spectra at low pressures, depending upon the relative depth of any Ti-bearing clouds. In general, the total cloud mass for Ti-bearing clouds is relatively low (see Table\,\ref{table:cloud_mass}) as Ti is the limiting element for these condensates. This suggests that if condensation occurs at low pressures and high altitudes, the cloud could be optically thin enough that molecular features, such as Na and K in the optical, could still be observed in the planetary transmission spectrum, assuming no other obscuring species is present. However, this would also reduce the liklihood of observing TiO/VO spectral features and thermal inversions would not be expected. 

The most likely species to form in super-hot Jupiter atmospheres at obscuring pressures ($\sim$mbar, see Fig. \ref{fig:shj_pt_profiles}) for transmission spectral observations are Al-bearing species. As before, due to the complex nature and interplay of Ca-Al-O condensates we treat them together when considering their effect on the atmospheric properties. The cloud mass of Al-bearing species is over ten times that of Ti-bearing species (Table \ref{table:cloud_mass}), due to the relative abundances of elemental material available (namely, Al vs. Ti), which will likely form massive clouds at altitudes where molecular features will be muted or obscured. 

Following the condensation of the high-temperature Al- and Ti-bearing condensates, the remaining atmospheric Ca is effectively removed by the subsequent formation of lower-temperature Ca-bearing condensates, such as gehlenite and akermanite (see \citealt{lodders2002}), greatly reducing the abundance of Ca in the upper atmosphere. Following the condensation of Ca/Al/Ti species, the most abundant condensible species available are Fe/Mg/Si which begin to condense near $\sim$1680\,K (at 0.1\,bar). At solar abundances these rock-forming elements form large cloud masses (see Table\,\ref{table:cloud_mass}) resulting in optically thick cloud layers. In an equilibrium scenario, these condensates are likewise expected to effectively remove Fe, Mg, and Si from the atmosphere above the Fe and Mg-silicate cloud layers (see \citealt{visscher2010}).
\begin{figure} 
\begin{center}
  \includegraphics[width=0.48\textwidth]{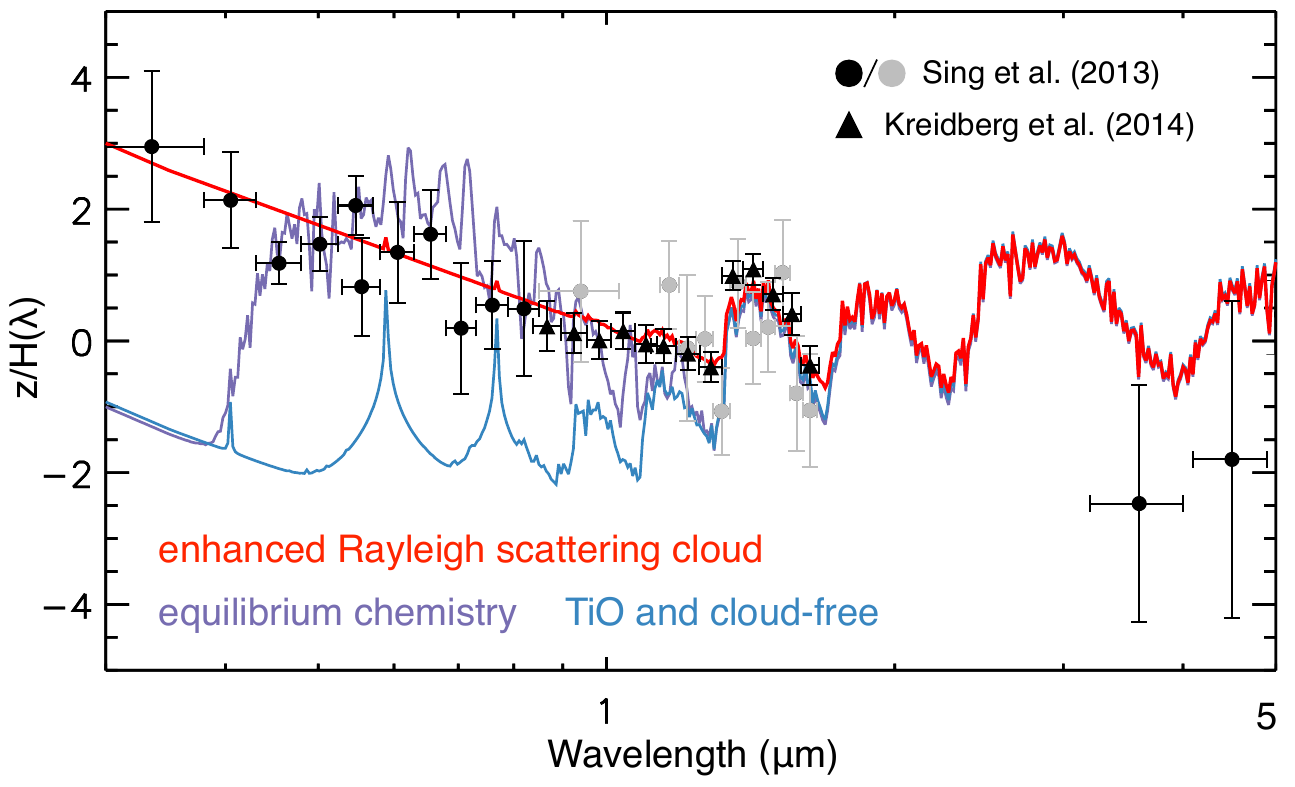}%w12_transmission_v2
\caption{Transmission spectrum of WASP-12b from HST STIS and \textit{Spitzer} observations (\citealt{Sing2016}, circles), and HST WFC3 (\citealt{kreidberg2015}, triangles). The gray points show the single transit observation results from HST WFC3 presented in \citealt{sing2013}, which are replaced here with the multiple transit observations from \citealt{kreidberg2015}. We show three isothermal models for the different atmospheric scenarios discussed in the text (\citealt{fortney2010})}\label{fig:shj_w12_transmission}
\end{center}
\end{figure}% I am a god among figures am I not :D

\subsection{The curious case of WASP-12b}\label{sec:w12}
WASP-12b orbits a metal-rich star with [Fe/H] = 0.3. The equilibrium temperature of WASP-12b, T$_{eq}\approx$2500\,K, suggests that condensation will not occur in the atmosphere and that TiO would be seen in the gas phase in a majority of the atmosphere. However, transmission observations have shown it to be consistent with a flat grey absorber extending into the near-IR with no detection of TiO or metal-hydrides (\citealt{sing2013}). Observation conducted by \citet{kreidberg2015}, which combined three transit observations in the G102 bandpass between 0.82--1.12\,$\mu$m and three transit observations in the G141 bandpass between 1.1--1.7\,$\mu$m, revealed a muted but present high-precision H$_2$O absorption feature centered at 1.4\,$\mu$m (see Fig. \ref{fig:shj_w12_transmission}). Under the assumption of chemical equilibrium the C/O ratio was constrained to less than one at a greater than 3$\sigma$ confidence. Additionally, emission spectra observations (\citealt{crossfield2012a}) lack a strong thermal inversion also indicative of a lack of TiO. Each observation is consistent with a solar C/O ratio (\citealt{madhusudhan2012}), and the presence of obscuring material in the atmosphere of WASP-12b. 

\begin{figure*} 
\begin{center}
  \includegraphics[width=0.97\textwidth]{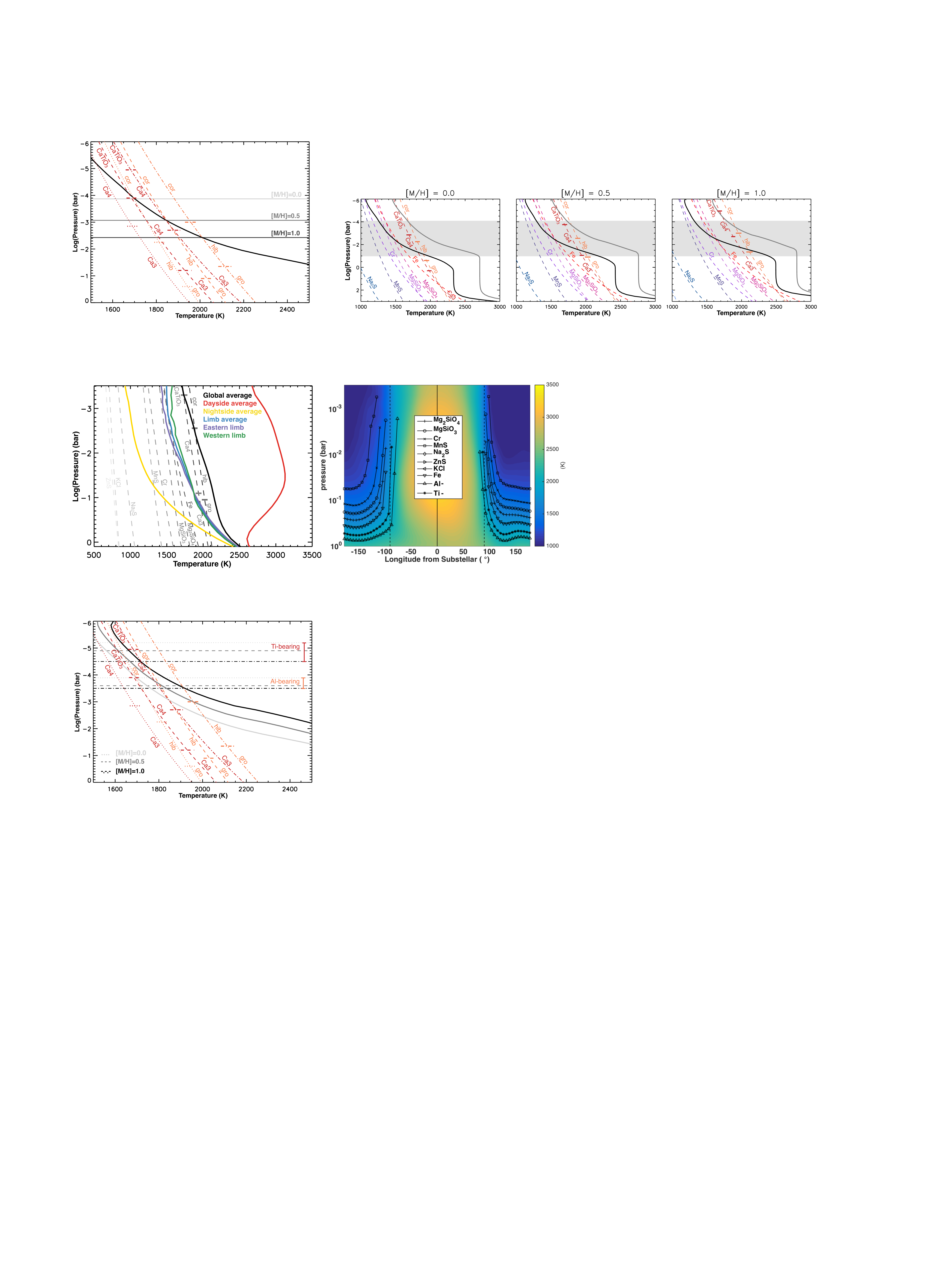}%W12_tp_tiff3
\caption{left: T-P profiles of WASP-12 showing global-, dayside-, nightside-, limb-, east terminator-, and west terminator-averaged following the calculations outlined in \citet{kataria2016}. Condensation curves are plotted in gray dashed lines. Right: Latitudinally-averaged temperature as a function of pressure and longitude from the substellar longitude. Overplotted are the condensation curves of multiple species demonstrating the regions where condensation should occur. Al- label refers to the condensates cor, hib, gro, and Ti- to CaTiO$_3$, ca4, ca3 (abbreviations same as Fig.\ref{fig:shj_pt_profiles}). The dashed vertical lines indicate the limbs of the planet observed globally in transmission spectra.}\label{fig:shj_w12_tp}
\end{center}
\end{figure*}
Comparison between theoretical spectra and all transmission measurements with HST and \textit{Spitzer} point strongly to abundant small particle aerosols in the atmosphere of WASP-12b \corr{(\citealt{sing2013,Sing2016})}. Figure\,\ref{fig:shj_w12_transmission} shows the two observations of WASP-12b over-plotted to demonstrate the consistency of the low S/N observations with the high-precision near-IR observations. We also show 1D isothermal atmospheric models with T$_{eq}$=2000\,K from a suite of models formulated by \citet{fortney2010} for a cloud-free atmosphere with equilibrium chemistry including TiO gas and equilibrium without TiO gas, with enhanced Rayleigh scattering. If TiO is depleted in the atmosphere via condensation, strong alkali metal absorption features (Na and K) are expected to become prominent in the optical. The lack of atomic or molecular absorption suggests the presence of a uniform optically thick absorber extending into the IR that mutes or obscures its gaseous spectral signatures. The best fitting 1D model for WASP-12b favors enhanced (100x) Rayleigh scattering in the optical extending into the near-IR covering H$_2$O absorption at 1.2\,$\mu$m and muting further absorption at 1.4\,$\mu$m. The presence of a strong Rayleigh-like slope in transmission towards the IR is indicative of a high altitude cloud composed of small particles. 

To estimate the possible cloud absorbers in the atmosphere of WASP-12b, we calculate the global temperature structure using the SPARC/MITgcm. The SPARC/MITgcm couples the MITgcm, a finite-volume code that solves the three-dimensional primitive equations on a staggered Arakawa C grid (\citealt{adcroft2004}) with a radiative transfer code that is a two-stream adaptation of a multi-stream radiative transfer code for solar system planets by \citet{Marley1999}. The radiative transfer code employs the correlated-k method with 11 bands optimized for accuracy and computational efficiency. The opacities are calculated assuming local thermodynamic and chemical equilibrium (\citealt{loddersfegley2002}). This code has been used extensively to model the atmospheric circulation of hot Jupiters, hot Neptunes and super Earths (e.g. \citealt{kataria2014}, \citeyear{kataria2015}, \citeyear{kataria2016}, \citealt{lewis2010}).These cloud-free models utilize the system parameters of \citet{chan2011}, and we compare our high temperature condensation curves to these computed three-dimensional T-P profiles. Figure\,\ref{fig:shj_w12_tp}a shows the T-P profiles computed across different regions of the atmosphere, where all-but the dayside-averaged profile cross significant condensation curves. 

To predict the regions of the atmosphere where condensation is expected to occur we map the latitudinally averaged temperature as a function of the pressure and longitude from the substellar longitude as derived from the GCM results and overplot the condensation curves (Fig.\,\ref{fig:shj_w12_tp}b). This comparison suggests that condensation is expected to occur on the limb of the planet where transmission spectral signatures are observed with both Al- and Ti-bearing species likely contributing to an obscuring cloud, and/or gas phase depletion. Given a relative cloud mass from solar composition (see Table\,\ref{table:cloud_mass}) and an atmospheric scale height of 950\,km, we calculate the spherical shell volume of the planetary atmosphere and assume all of our cloud material is condensed into one scale height. We approximate the optical depth for a cloud composed of corundum at 1\,mbar (Fig.\ref{fig:shj_w12_tp}) in the atmosphere of WASP-12b to be, $\tau_{Al_2O_3}$$\approx$\,0.76, given a particle size of 0.1\,$\mu$m at $\lambda$=1.4$\mu$m. We follow the same calculation for hibonite at 0.1bar (Table\ref{table:cloud_mass}, Fig.\ref{fig:shj_w12_tp}) which gives $\tau_{CaAl_{12}O_{19}}\approx$\,0.82. This shows that if condensation of Al-bearing species occurs the cloud will be optically thick enough to effectively obscure molecular features in the transmission spectrum of WASP-12b. 

To determine the estimated cloud particle sizes and impact of cloud species on the observed transmission spectrum we calculate the transmission spectrum for Al-bearing and Ti-bearing condensates using Mie theory (\citealt{bohren1983}) and following the methods outlined in \citet{wakeford2015}. As shown in \citet{wakeford2015}, it would be difficult to distinguish exactly which Al-bearing, or Ti-bearing condensate is present and therefore we use a single example for each; adopting optical properties available in the literature for corundum (Al$_2$O$_3$, \citealt{koike1981}), and perovskite (CaTiO$_3$, \citealt{posch2003b}). For each of the condensates considered we use the temperature and pressure where the condensation curve crosses the global-averaged T-P profile (see Fig.\,\ref{fig:shj_w12_tp}a). While the cloud will likely extend over a wider region of the atmosphere, we use this as a first approximation to demonstrate the scattering effects of the condensates. 
\begin{figure} 
\begin{center}
  \includegraphics[width=0.48\textwidth]{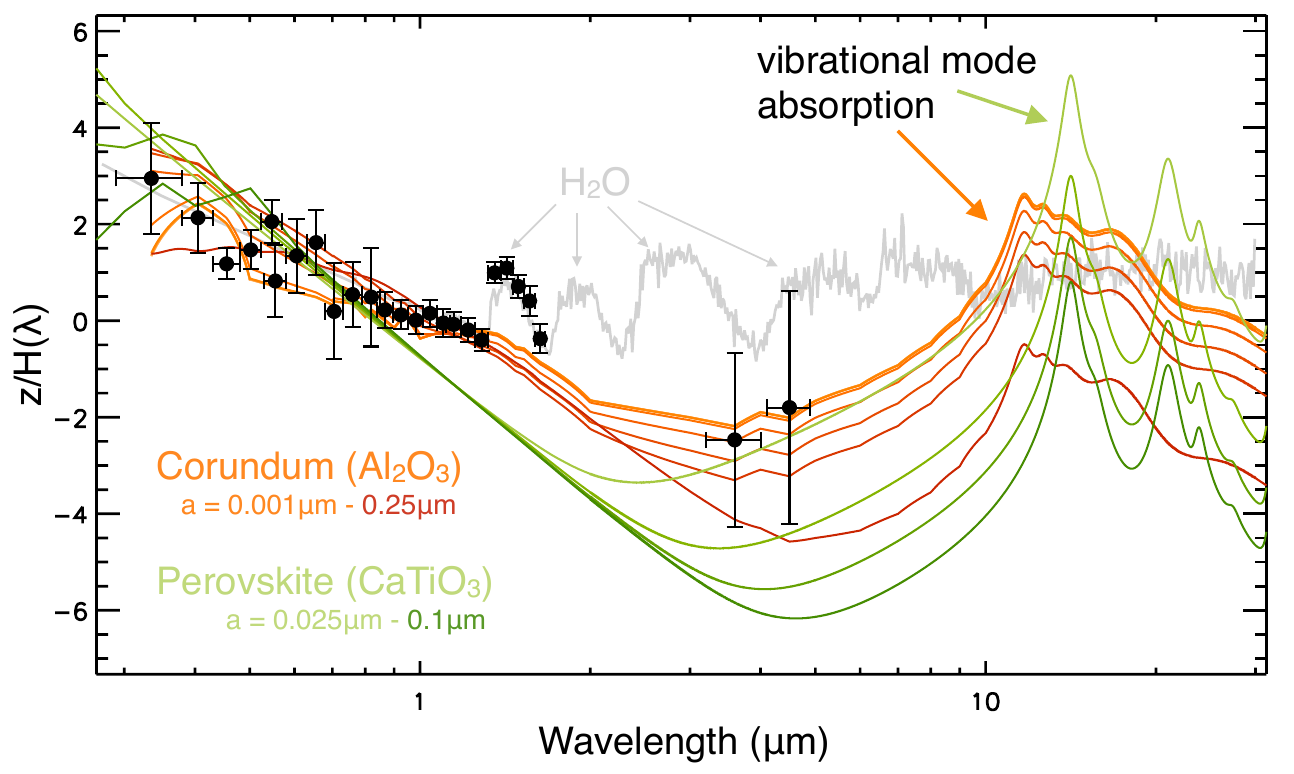}%w12_transmission_v2
\caption{Transmission spectrum of WASP-12b (black points) with the enhanced Rayleigh isothermal model (grey). The computed condensate transmission spectra of corundum (orange) and perovskite (green) are shown for the range of fitted particle sizes. We include the wavelength range covered by JWST which covers the expected vibrational-mode absorption features.}\label{fig:shj_w12_clouds}
\end{center}
\end{figure}% I see your scattering and I raise you a vibrational mode!

As shown in \citet{lecavelier2008} and \citet{wakeford2015} the transmission spectral properties of the cloud are defined by the maximum particle size ($a$) in a distribution, as the cross-section is proportional to $a^6$. Using a grid of \corr{single} particle sizes \corr{from 0.0001--2.5\,$\mu$m}, we compute the transmission spectrum for each species and fit them to the observations using least-squares minimization (\citealt{markwardt2009}). From this fitting process (outlined in \citealt{wakeford2015}) we determine that the best fit particle sizes, defined by the reduced $\chi^2$, for corundum to be 0.001--0.25\,$\mu$m and 0.025--0.1\,$\mu$m for perovskite. Figure\,\ref{fig:shj_w12_clouds} shows the computed transmission spectra for corundum and perovskite for the range of the best fitting maximum particle size in the cloud. Note, the condensate transmission spectra show the scattering properties of the corundum and perovskite separately without the inclusion of additional molecular absorbers. The condensate transmission spectra show the best fit to the data, including the \textit{Spitzer} points, where previous models have struggled (Fig.\,\ref{fig:shj_w12_clouds}). The major vibrational mode of the Al-O bond in a majority of the Al-bearing species will dominate the absorption in the transmission spectra of super-hot Jupiters between 9--28\,$\mu$m. Given the vibrational mode structure of Ti-bearing condensates the Ti-O bond will generate absorption features between 16--66\,$\mu$m, with perovskite having a broad double peak structure at 16 and 21\,$\mu$m (\citealt{wakeford2015}). \corr{Given the slant geometry through the planetary atmosphere observed during transit, minor condensates with low optical depths can become significant absorbers compared to when viewed normal to the planet (\citealt{fortney2005}). Therefore the vibrational mode signatures may become prominent in exoplanet spectra and potentially become stronger than those detected in cloudy brown dwarf atmospheres where small particles are seem to produce stronger Si-O features (e.g. \citealt{cushing2006,riaz2009}).} 

Importantly, JWST will provide key wavelength coverage over these absorption features, and if present potentially distinguish between obscuring Al-bearing clouds covering TiO features, and/or Ti-bearing clouds reducing the TiO gas opacity. Additionally, observations with HST in the optical are still vital to estimate cloud particle sizes as a predictor of vibrational mode absorption features in the IR.

% 6
%CONCLUSIONS
%
\section{Conclusion}
In this paper we explore the high temperature condensation sequence relevant to super-hot (T$_{eq}$\textgreater1800\,K) Jupiter exoplanet atmospheres. We present new condensation curves of the highest temperature Al-, and Ti-bearing condensates for an improved condensation sequence. Given the lack of temperature inversions seen in hot Jupiters (e.g. \citealt{line2014}) and lack of TiO absorption features in optical spectra (\citealt{Sing2016}), our models here suggest that super hot Jupiter T-P profiles can naturally cross condensation curves of several refractory condensates. We expect these planets to form substantial cloud decks (in the correct composition, temperature, gravity phase space). We discuss the impact of increased metallicity on the cloud mass, and demonstrate that increasing the abundance of metals available results in a higher cloud opacity as more mass is available to be condensed. Increased metallicity also pushes the point of condensation in a planetary T-P profile to deeper layers of the planetary atmosphere increasing the density of the cloud as the spherical shell volume decreases. \corr{This has the greatest impact in profiles where the condensation point is pushed into observable regions of the atmosphere and/or where the profile is near isothermal such that the spherical scale volume is decreased with little change to the planetary scale height.}

In the super-hot regime cloud species are dominated by Al, and Ti, where Al-bearing elements are most abundant and Ti is the limiting element. Using WASP-12b as an example we show that TiO will likely be reduced by condensation or obscured by Al-bearing clouds when temperatures in any part of the atmosphere around the limb fall below $\approx$1900\,K, and will therefore not be seen as a major gas phase absorber in the optical transmission spectrum. Following Mie theory the shape of the transmission spectrum in the optical indicates the maximum size of particles forming a cloud. Flat spectra imply large uniform scatterers which will likely obscure absorption features into the IR, while a Rayleigh-like slope is indicative of small sub-micron sized particles. We suggest, given the vibrational mode absorption from the condensates for clouds made of small sub-micron particles compositional distinctions between Al and Ti species can be potentially detected at JWST wavelengths.

%Grrrrrrrr Arrrrrg!

% 6
%ACKNOWLEDGEMENTS
%
\section{Acknowledgments}
H.R. Wakeford acknowledges support by an appointment to the NASA Postdoctoral Program at Goddard Space Flight Center, administered by ORAU and USRA through a contract with NASA. C.V. acknowledges support by NSF grant AST-1312305. The authors would also like to thank the annonomous referee for their comments and suggestions. 

%% in parentheses, after the keyword, but they will not be verified.

%{\it Facilities:} \facility{HST (WFC3)}. \\

% 6
%BIBLIOGRAPHY 
%
\bibliographystyle{mn2e}
\bibliography{references}

\label{lastpage}
% Is that it, are we done now?
\end{document}